\documentclass{article}
\usepackage[utf8]{inputenc}

\usepackage{graphicx}
\graphicspath{ {./img/} }

\usepackage{fixltx2e}
\usepackage{dcolumn}
\usepackage{array} \newsavebox\cellbox
\usepackage{multirow}
\usepackage{amsmath}

\usepackage[font=small]{caption}

\usepackage{booktabs}

\makeatletter
\newcommand*{\addFileDependency}[1]{
  \typeout{(#1)}
  \@addtofilelist{#1}
  \IfFileExists{#1}{}{\typeout{No file #1.}}
}
\makeatother

\newcommand*{\myexternaldocument}[1]{%
    \externaldocument[supp-]{#1}%
    \addFileDependency{#1.tex}%
    \addFileDependency{#1.aux}%
}

\usepackage{xcolor}

\usepackage{xr}
\myexternaldocument{SI}

\usepackage[capitalise]{cleveref}
\usepackage{authblk}

\usepackage{listings}

\definecolor{mygreen}{rgb}{0,0.6,0}
\definecolor{mygray}{rgb}{0.5,0.5,0.5}
\definecolor{mymauve}{rgb}{0.58,0,0.82}

\usepackage{biblatex}
\title{Variable Markov dynamics as a multi-focal lens to map multi-scale complex networks}
\author[*,1,2,3]{Daniel Edler}
\author[1,4]{Jelena Smiljanić}
\author[1]{Anton Holmgren}
\author[2,3,5,6]{Alexandre Antonelli}
\author[1]{Martin Rosvall}

\affil[1]{{\small Integrated Science Lab, Department of Physics, Ume\aa\ University, Ume\aa, Sweden}}
\affil[2]{{\small Department of Biological and Environmental Sciences, University of Gothenburg, Gothenburg, Sweden}}
\affil[3]{{\small Gothenburg Global Biodiversity Centre, Department of Biological and Environmental Sciences, University of Gothenburg, Gothenburg, Sweden}}
\affil[4]{{\small Institute of Physics, University of Belgrade, Pregrevica 118, Belgrade, Serbia}}
\affil[5]{{\small Department of Plant Sciences, University of Oxford, Oxford, United Kingdom}}
\affil[6]{{\small Royal Botanic Gardens, Kew, Richmond, Surrey, United Kingdom}}
\affil[*]{{\small Corresponding author: daniel.edler@umu.se}}

\date{November 2022}

\begin{document}

\maketitle

\begin{abstract}
From traffic flows on road networks to electrical signals in brain networks, many real-world networks contain modular structures of different sizes and densities. 
In the networks where modular structures emerge due to coupling between nodes with similar dynamical functions, we can identify them using flow-based community detection methods. 
However, these methods implicitly assume that communities are dense or clique-like which can shatter sparse communities due to a field-of-view limit inherent in one-step dynamics. 
Taking multiple steps with shorter or longer Markov time enables us to effectively zoom in or out to capture small or long-range communities. However, zooming out to avoid the field-of-view limit comes at the expense of introducing or increasing a lower resolution limit.
Here we relax the constant Markov time constraint and introduce variable Markov dynamics as a multi-focal lens to capture functional communities in networks with a higher range of scales. With variable Markov time, a random walker can keep one-step dynamics in dense areas to avoid the resolution limit and move faster in sparse areas to detect long-range modular structures and prevent the field-of-view limit.
We analyze the performance of variable Markov time using the flow-based community detection method called the map equation. We have implemented the map equation with variable Markov time in the search algorithm Infomap without any complexity overhead and tested its performance on synthetic and real-world networks from different domains. Results show that it outperforms the standard map equation in networks with constrained structures and locally sparse regions. In addition, the method estimates the optimal Markov time and avoids parameter tuning.
\end{abstract}

\section{Introduction}
Good maps simplify and highlight important structures in complex data. Applied to networks, we are interested in describing community structures that give a coarse-grained representation of the system. Community structures can identify important functional modules, for example, genes with similar expression patterns in gene co-expression networks or research areas and disciplinary boundaries from scientific citation networks~\cite{poldrack_long-term_2015,rosvall_memory_2014}. In these and similar examples where modular structure determines network dynamics, flow-based community detection methods are beneficial in identifying groups of nodes with similar dynamical function~\cite{rosvall_maps_2008}. These methods use flow to model a dynamic process and highlight groups of nodes where flow remains relatively long.

However, functional modules may have a structure not typically recognized as a community by flow-based community detection methods that assume clique-like communities~\cite{schaub_encoding_2012}. For example, longer neural pathways have an important function in connecting denser subnetworks of gray matter in the brain, but network structures that require many steps to traverse tend to be subdivided into multiple communities, creating a disconnect between structure and function for some real-world networks~\cite{schaub_markov_2012}.

The map equation is a dynamic approach to community detection that models network flows by a random walker and optimizes communities to capture areas where the random walker stays relatively long~\cite{rosvall_maps_2008, edler_infomap_2017}.
It describes the modular map by encoding the random walk with a multilevel codebook quantifying the description length of movements within and between communities.
Conceptually, this approach works well to find modular regularities in networks such as standard benchmark networks where nodes inside a community have high internal link density and lower link density between communities~\cite{lancichinetti_community_2009}. However, real-world networks with structures that deviate from this assumption require modification of flow models.

In contrast to other popular community detection methods, such as modularity or the stochastic block model, the map equation effectively eliminates the resolution limit for networks with relatively small communities~\cite{fortunato_resolution_2007,peixoto_parsimonious_2013,kawamoto_estimating_2015}.
However, community detection methods based on one-step random walks have a field-of-view limit -- an upper limit in the effective diameter of the communities they can detect~\cite{schaub_encoding_2012}. The random walker will remain within subgroups of nodes inside large, sparse communities and break them into smaller communities, leading to overfitting. Changing the Markov time, the time between reporting the position of the random walker~\cite{kheirkhahzadeh_efficient_2016}, keeps larger structures from breaking apart into multiple communities. However, increasing the global Markov time to avoid the field-of-view limit comes at the expense of losing finer resolution in denser areas due to a forced resolution limit. A single (fixed) time cannot resolve a highly heterogeneous community structure~\cite{schaub_markov_2012}.

Here we introduce variable Markov time that relaxes the constraint of a global resolution parameter that otherwise forces us to choose between detecting large-scale structures above the field-of-view limit or highlighting small dense structures. With variable Markov time, a random walker moves faster on sparse areas to avoid the field-of-view limit without generating a resolution limit on denser areas. Increasing the random walker's dynamic range enables us to capture relevant modular flow patterns with a wider range of scales within a single map.

We analyze synthetic and real-world networks with constrained or locally sparse substructures and compare with two popular algorithms -- the modularity-based Leiden method~\cite{traag_louvain_2019} and the degree-corrected stochastic block model~\cite{peixoto_merge-split_2020} (DC-SBM). A toy network with cliques connected by chains of different lengths highlights challenges in the methods. While the performance of the map equation and Leiden algorithms depend on the balance between chain length and clique size, the performance of the DC-SBM method depends on the size of the cliques. Results for synthetic and real-world networks show that the map equation with variable Markov time reduces over-partition in sparse network regions without imposing a resolution limit.

\section{Method}
To not over-partition sparse community structures we need to increase the Markov time. To not under-partition dense structures, we need to keep the Markov time at one, corresponding to one-step random walks. To solve these conflicting requirements, we relax the constraint of a global Markov time and introduce a variable Markov time.

A global change of Markov time can be approximated for the map equation by rescaling the link flow by the Markov time~\cite{kheirkhahzadeh_efficient_2016}. For variable Markov time, we need to define a local Markov time that increases with sparsity. A local measure of sparsity is the number of connected links to a node: the node's degree. However, a related property more relevant from a dynamic community detection perspective is the random walker's visits frequencies, i.e., the flow. This is calculated using the PageRank algorithm~\cite{page_pagerank_1999} modified for unrecorded teleportation to links~\cite{lambiotte_ranking_2012}. For an undirected unweighted network, the visit frequency, or flow, \(p_i\) of node \(i\) is simply the degree \(k_i\) divided by total degree \(k_{tot}\). For weighted networks, the degree is replaced by the weighted degree. For directed networks, nodes have separate in- and out-degrees that decouple enter and exit probabilities of communities, and visit frequencies are no longer a simple function of the local degrees. Low-degree nodes may channel large one-way flow, and high-degree nodes in communities with low enter and high exit probability receive low flow.

To even out the flow between nodes, a local Markov time would have to be inversely proportional to the local flow. To not incur a resolution limit we use the densest part of the network as a baseline to keep Markov time 1 as a minimum. This gives us a variable Markov time defined as \( p_{max} / p_i \) where \(p_{max}\) is the flow of the most visited node. For undirected unweighted networks, this Markov time becomes \(k_{max}/k_i\). As the link flow describes transition rates for the random walker, and the Markov time inflates link flow without increasing node flow, variable Markov time describes variable encoding rates, i.e., variable number of steps between recording the node visits. In an undirected unweighted network where a hub with degree eight sets the baseline for Markov time 1, the local Markov time levels out the link flow such that  the incoming and outgoing flow is eight for all nodes. For a chain with degree two, this means that the random walker moves four times as often through it but only records the node visits every four steps on the chain. This model for variable Markov time describes a random walker that moves with a constant flow rate rather than a constant speed or step rate.

However, many real-world networks have a fat-tailed degree distribution, where most nodes have only a few links and a few nodes have degrees that greatly exceeds the average, such as the most popular accounts or viral content in social networks. In that case, the local Markov time for most nodes can be very high with an uncertainty covering orders of magnitude. As the optimal number of communities depends on the flow between communities, the optimal modular solution with variable Markov time defined above is not robust for networks with fat-tailed degree distributions such as power-law distribution. To increase the robustness of such networks, we can scale the baseline and local flow logarithmically. That is, instead of doubling the local Markov time on a chain in an undirected unweighted network for each doubling of the maximum degree, it will, using a logarithm with base two, be increased by one each time. By keeping variable Markov time as a ratio between the maximum scale and a local scale, for the logarithmic scaling to work, we need to work with the unnormalized flow \(p_i k_{tot}\) which for undirected networks are the node degrees. We also need a minimum scale to avoid the zero division problem when we have dangling or disconnected nodes. As the sparsest structure that can be broken due to a field-of-view limit is a chain, where each node has degree two, we use that as a lower limit. This model for variable Markov time describes for an undirected unweighted network a random walker that moves with a constant entropy rate rather than a constant speed or step rate.

To support a broad spectrum of networks, we define a damping parameter \(\zeta\) to scale the local Markov time inversely proportional to, for undirected networks, the degree and the logarithm of the degree for damping zero and one respectively. We can then define a local Markov time \(t_i^\zeta\) for each node as
\[
t_i^0 = \frac{p_{max}k_{tot}}{\min(1, p_i k_{tot})}
\]
and
\[
t_i^1 = \frac{\log_2 p_{max}k_{tot}}{\min(1, \log_2 p_i k_{tot})}.
\]
As Markov time rescales the link flow, Markov time on directed links is the Markov time for the source nodes defined above. For undirected links, to keep minimum Markov time \(1\), we define the Markov time between node \(i\) and \(j\) as
\[
t_{ij} = \min(t_i, t_j).
\]

To handle a broader range of networks, we use damping one if no other value is specified. For constrained networks with more even degree distribution, we might need damping zero to avoid the field-of-view limit. To explore the effect of variable Markov time, we use the Tsallis entropy property that for a uniform probability distribution of \(k\) categories, it interpolates from linear to logarithmic scaling of \(k\). This means that we use the entropic index of the Tsallis entropy as a damping parameter, which supports any real value \(\zeta \geq 0\). Higher damping reduces variability with the limit
\[\lim\limits_{\zeta \to \infty} t_i^\zeta = 1.\]

For details on the damping parameter, see Supporting Information.
We have implemented variable Markov time in the map-equation-based network clustering algorithm Infomap\cite{Edler_The_MapEquation_software_2022} with no computational complexity overhead (see SI).

\section{Results}
We evaluate community detection with variable Markov time using Infomap and compare our results with both standard Infomap and two other common community detection methods based on modularity and the stochastic block model.
We test the method on synthetic and real-world networks with multi-scale structures.

\subsection*{Synthetic cliques-and-chain network}

\begin{figure}[htbp]
  \centering
  \includegraphics[width=\textwidth]{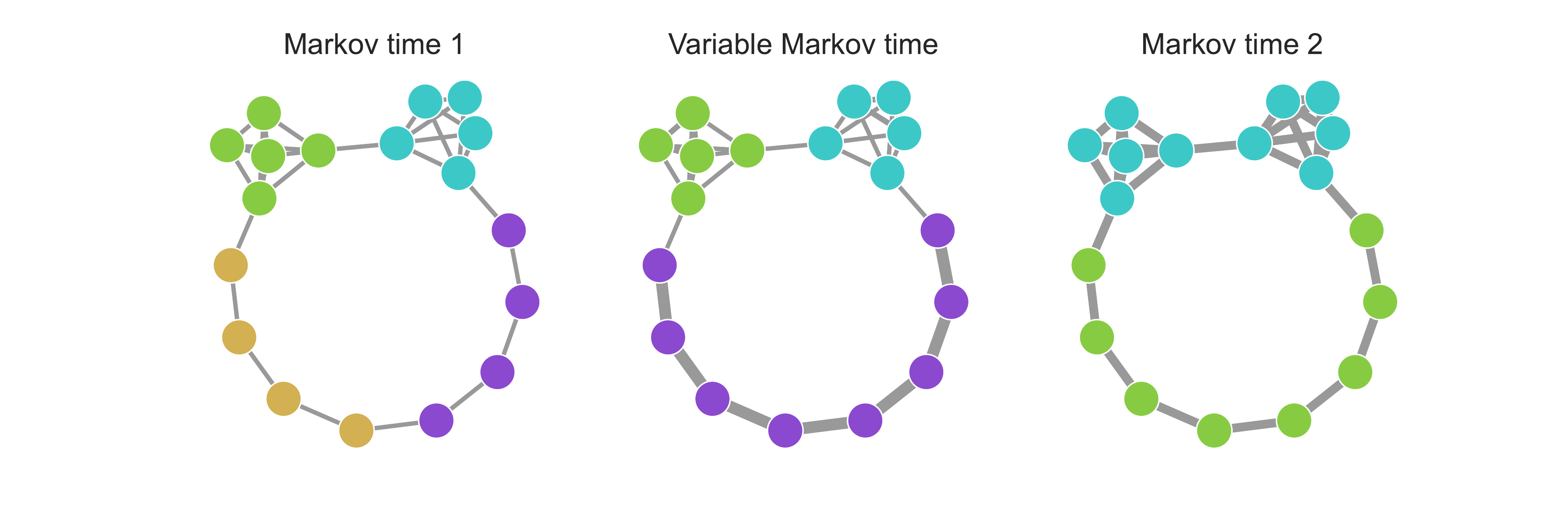}
  \caption{Constant and variable Markov time in a toy network. Colors show detected communities, which capture nodes where a random walker tends to spend a relatively long time before exiting. Markov time describes the number of steps before reporting the node. With a default one-step random walk, the chain is optimally partitioned into two communities. With Markov time 2, the random walker takes two steps per recording which double the flow between nodes, indicated by the wider links. This increases the field-of-view limit and enables the random walker to optimally explore the chain as a single community. However, this comes with the cost of an increased resolution limit, which makes it harder to detect modular patterns in denser areas. Variable Markov time relaxes the constraint of moving with a constant encoding rate and adapts the encoding rate to the level of sparsity. Variable Markov time increases the gap between the lower resolution limit and upper field-of-view limit and enables us to detect community structures with a broader range of scales than was previously possible.}
  \label{fig:toy-schematic}
\end{figure}
Our benchmark network composed of two cliques and a chain (Fig.~\ref{fig:toy-schematic}) highlights both the field-of-view limit -- an upper scale on the size of a network structure that can be detected -- and the resolution limit -- a lower scale on the size of a network structure that can be detected. We define these limits in terms of the global Markov time: the field-of-view limit sets a minimum Markov time needed to detect the chain structure as one community and the resolution limit a maximum Markov time to detect two clique structures as separate. A Markov time between those limits makes it possible to detect all three parts of the benchmark network~(\cref{supp-fig:fov} in SI).


\begin{figure}[htbp]
  \centering
  \includegraphics[width=\textwidth]{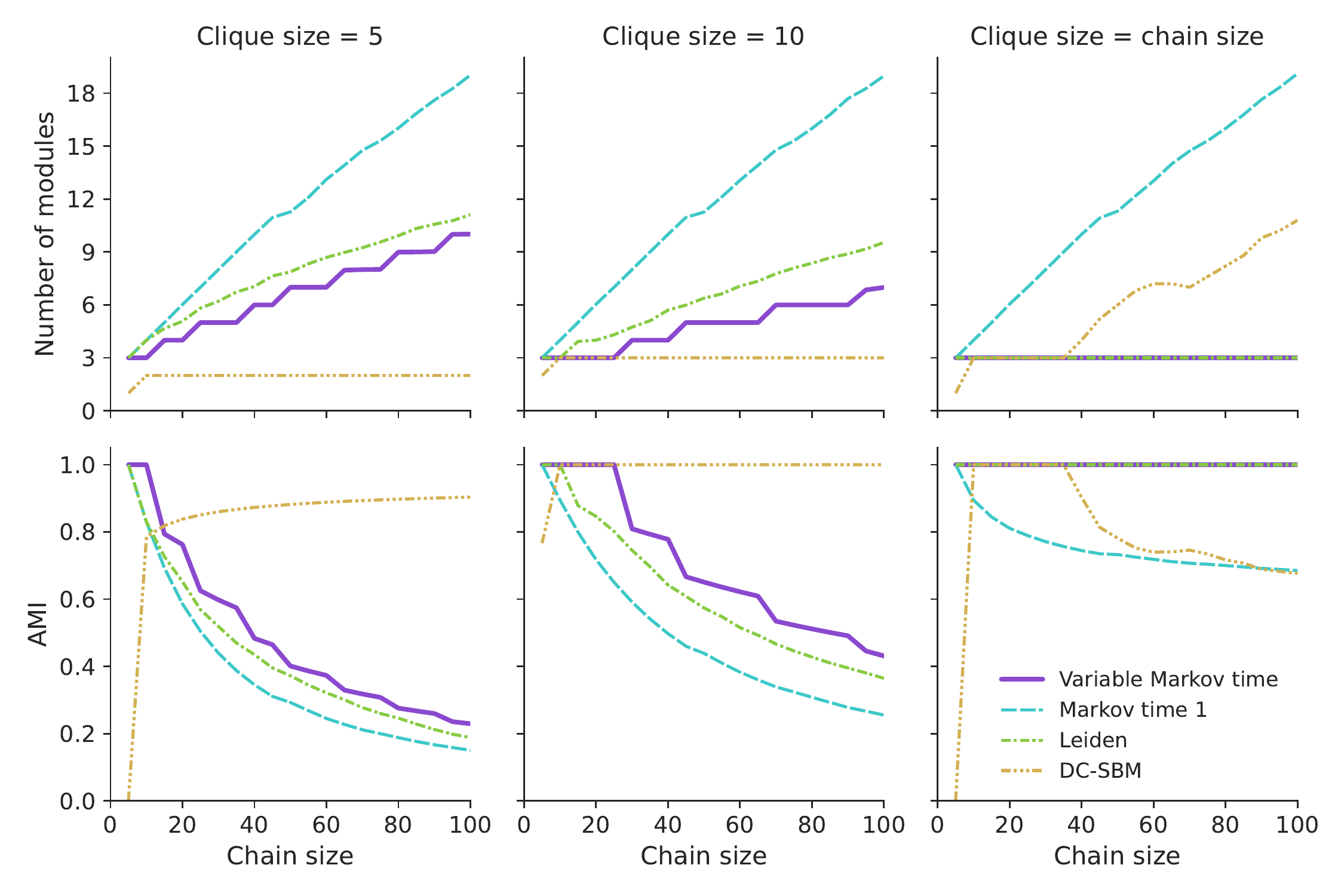}
  \caption{Mean number of communities (top row) and adjusted mutual information (bottom row) in synthetic networks consisting of two equally-sized cliques connected by a chain that represent planted communities. Results are obtained by the map equation with variable Markov time and with Markov time 1, the Leiden method, and the degree-corrected stochastic block model (DC-SBM). Results are averaged over 100 algorithm searches. }
  \label{fig:toy}
\end{figure}

We analyze the field-of-view limit by varying chain size (Fig.~\ref{fig:toy}). The standard map equation with Markov time 1 detects 3 communities for small chains with 5 nodes. As the number of nodes in the chain increases, the field-of-view limit forces the map equation to split the chain into smaller communities. We consider networks where the number of nodes in the clique is 5, 10, or equal to the number of nodes in the chain and use adjusted mutual information (AMI) to measure similarity between detected and planted communities~\cite{vinh2010ami}. The results show that the clique size does not affect the number of communities the standard map equation detects. However, the AMI score is higher for networks with larger clique-like communities. With variable Markov time, the field-of-view limit is pushed further away with increasing clique size and disappears completely when the clique size grows with the chain size. In this case, the resolution limit also never drops. This means that the cliques and the chain can be detected scale-independent.

We compare the performance of the map equation to the Leiden and the DC-SBM methods. The Leiden method shows a similar trend in detecting a higher number of communities with increasing chain size in networks with clique sizes 5 and 10. Compared to the map equation with variable Markov time, it starts to split the chain earlier. It recovers planted communities successfully in networks where clique sizes grow with chain sizes. Unlike the map equation and the Leiden method, the DC-SBM is more sensitive to the clique size than to the chain size. It successfully recovers planted communities in the networks where cliques have between 10 and 40 nodes. In a network with small cliques with 5 nodes, it underfits and detects two communities, while in networks with 40 or more nodes it overfits.

\subsection*{Infrastructure networks}

\begin{table}[!h]
\centering
\caption{Comparison between partitions detected by the map equation with Markov time 1 (MT=1) and variable Markov time (VMT) for four real-world infrastructure networks. For each network, we run 500 Infomap optimization trials. The columns $m$ denote the average number of communities detected by the map equation with Markov time 1 and variable Markov time, respectively. The columns $R$ report the average AMI between pairs of modular partitions, while the columns CV$(L)$ report the average coefficient of variance of the codelengths~$L$.}
\label{table:realnet}
{\small
\begin{tabular}{Wl{1.4cm}Wr{1.1cm}Wr{0.7cm}Wr{1.1cm}Wr{0.8cm}Wr{0.95cm}Wr{0.6cm}Wr{0.95cm}Wr{0.6cm}}
        &       &       & \multicolumn{2}{c}{Communities $m$} & \multicolumn{2}{c}{AMI $R$} & \multicolumn{2}{c}{CV$(L)$} \\ 
Network & Nodes & Links & MT=1 & VMT & MT=1 & VMT & MT=1 & VMT \\
\midrule
\rule{0pt}{3ex}Streets & 179 & 230 & 23.73 & 16.39 & 0.85 & 0.90 & 0.003 & 0.003 \\ \addlinespace
El. circuits & 122 & 189 & 18.43 & 7.66 & 0.86 & 0.94 & 0.002 & 0.003 \\ \addlinespace
Power grid & 2783 & 3762 & 245.67 & 75.95 & 0.88 & 0.90 & 0.001 & 0.002 \\ \addlinespace
Transport & 271 & 312 & 37.42 & 20.66 & 0.95 & 0.96 & 0.001 & 0.002 \\
\end{tabular}}
\end{table}
We consider four infrastructure networks as examples of real-world networks with constrained structures and locally sparse regions. The networks include the street network in Brasilia (Streets) \cite{streets}, electronic circuits network (El. circuits) \cite{elcircuits}, European power grid network (Power grid) \cite{schaub_markov_2012} and London transport network (Transport) \cite{transport}. All the networks are undirected. Due to their constrained structure, the degree distribution is relatively even, and we use variable Markov time with damping zero. Table~\ref{table:realnet} provides summary statistics for the performance of the map equation with Markov time 1 and variable Markov time in these networks. In all the networks, variable Markov time reduces the number of communities by more than $30$~percent. Since the networks contain locally sparse regions, we expect the map equation with Markov time 1 to break these subgroups of nodes into smaller communities, while variable Markov time should provide a more robust solution. To test for the robustness of the detected partitions, we run Infomap with 500 different seeds and calculate pairwise partition similarities using AMI. The low values of the codelength's coefficient of variance CV$(L)$ indicate that the partitions have similar quality. Except for the Transport network, the average AMI is below 0.9 for partitions found using Markov time 1 ($R_\textsubscript{MT=1}$). Identification of dissimilar partitions of similar quality suggests that the so-called solution landscape might be degenerate. This means that it is impossible to distinguish any optimal partition. Variable Markov time is a better model for these networks in the way that it prevents over-partitioning sparse regions and provides more robust solutions with $R_\textsubscript{VMT} \geq 0.9$ in all the networks.

\begin{figure}[htbp]
  \centering
  \includegraphics[width=\textwidth]{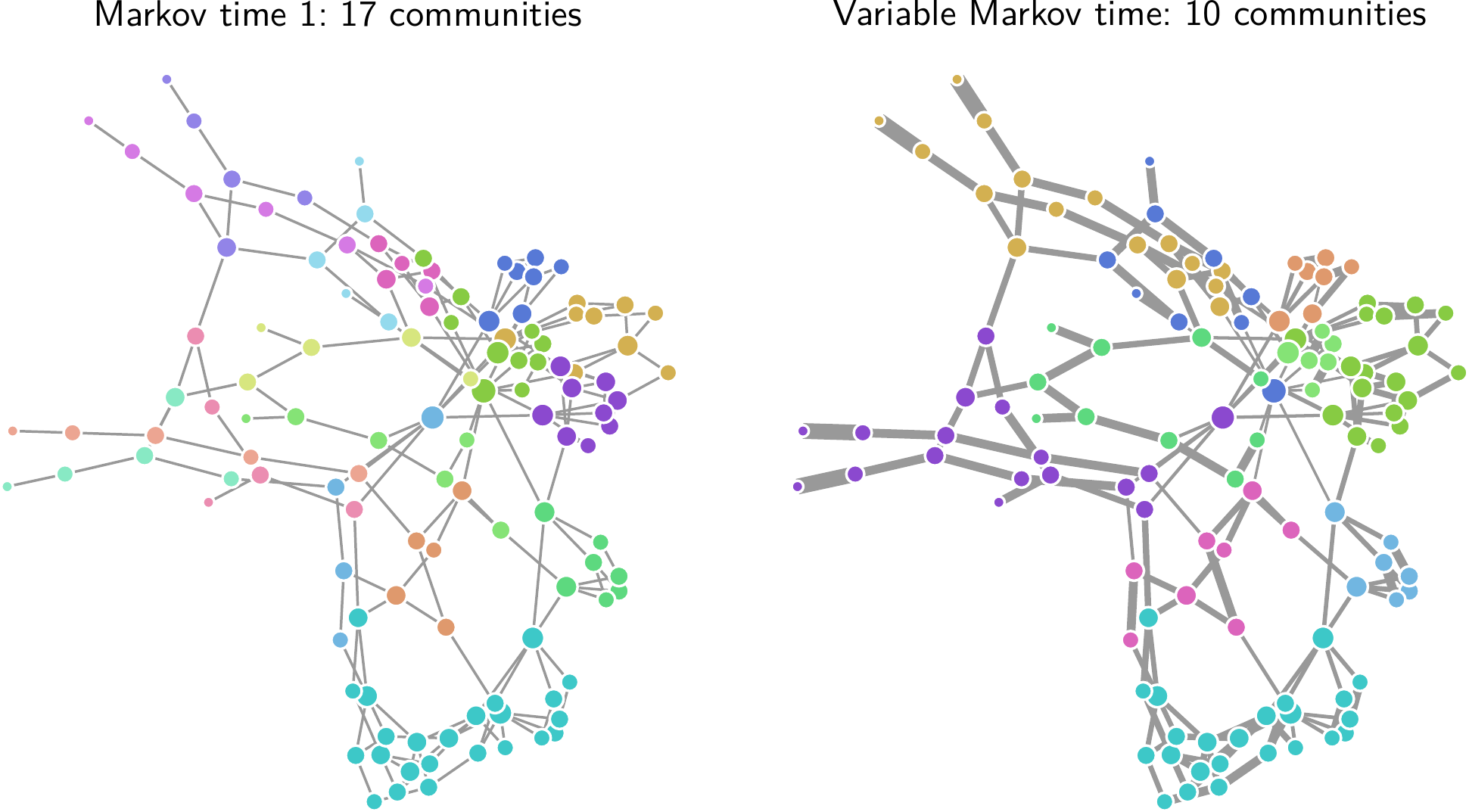}
  \caption{Electronic circuits network with 122 nodes and 189 links. Node colors indicate community assignments and Markov time-dependent link flow indicates the link thicknesses. With Markov time 1, the map equation returns 17 communities. Variable Markov time induces higher link flows in sparse regions and prevents splitting chain-like structures into separate communities, resulting in 10 communities.}
  \label{fig:elcircuits}
\end{figure}
To gain a better insight into the link flow distribution in sparse network regions and how it affects community detection, we illustrate it using the electronic circuits network (Fig.~\ref{fig:elcircuits}). Markov time 1 results in uniform link flow distribution, and the map equation prefers to split the groups of nodes with chain-like structures and large diameters resulting in 17 communities. Variable Markov time increases link flows in sparser regions, which enables the random walker to explore it faster resulting in sparser regions kept together. The number of communities is reduced from 17 to 10, and the average AMI increases from 0.86 to 0.94 (Table~\ref{table:realnet}).

\subsection*{Species occurrence network}
The second real-world network is a bipartite species occurrence network between taxonomic and geographical nodes, where each taxonomic node is connected to the geographical nodes where it occurs~\cite{edler_infomap_2017}. Here the occurrence data is the World Checklist of Vascular Plants (WCVP), a comprehensive, expert-curated dataset of harmonized taxonomic names and geographic distributions. It includes all currently accepted vascular plant species: flowering plants, conifers, ferns, clubmosses, and firmosses.~\cite{govaerts_world_2021, WCVP}. The geographical nodes are so-called ``botanical countries'', level 3 in the World Geographical Scheme for Recording Plant Distributions (WGSRPD) defined by the Taxonomic Databases for Plant Sciences (TDWG). It provides an agreed system of geographical units on approximately country level for most small countries, often with subdivisions for larger countries. It is the highest available resolution for recording plant distributions through presence and absence information from local floras and inventories~\cite{brummitt_world_2001,tietje_global_2022}.

\begin{figure}[htbp]
   \centering
   \includegraphics[width=\textwidth]{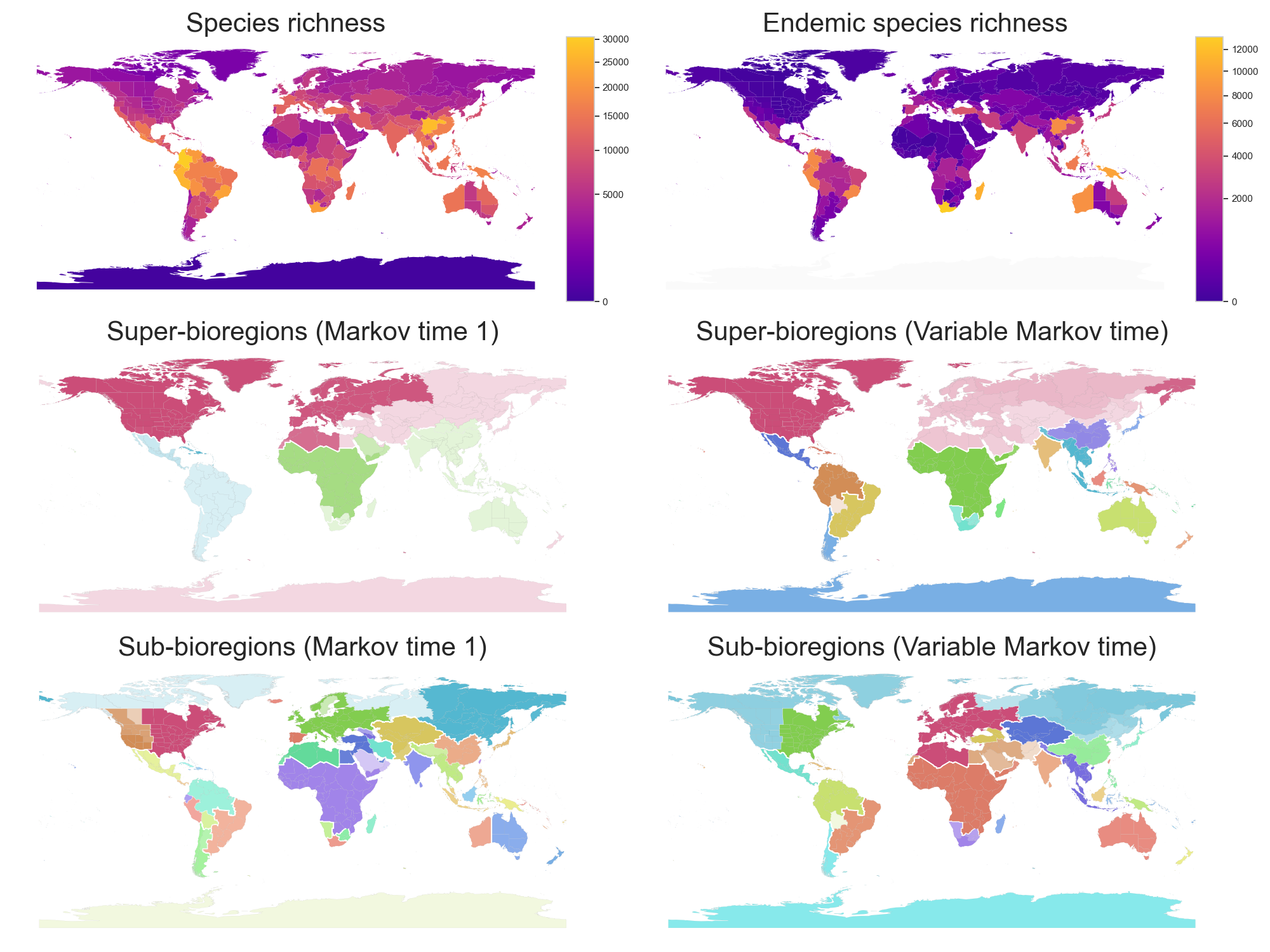}
   \caption{Species occurrence network. By connecting all currently accepted vascular plant species (WCVP) to all of the 368 ``botanical countries'' where they occur (WGSRPD), they form a network. We partitioned the network 1000 times using Infomap, with and without variable Markov time and discarded the worst percentile. With significance clustering, we find the subset of geographical nodes within each community that is together in at least 90 percent of the partitions, and the opacity indicates how often non-core nodes are partitioned together with the core. For the community boundaries, the thickness and opacity on the white line indicate how often the two geographical nodes on each side of the line are partitioned into different communities.}
   \label{fig:bioregions}
\end{figure}

The network contains 428,126 species nodes and 368 geographical nodes with a median of five thousand species each. Half of the species are endemic to their region, meaning that they occur only in a single botanical country, and 90~percent of the species occur in less than ten (Fig.~\ref{fig:bioregions}).

For species occurrence networks, a random walker alternates between exploring a random species at the current geographical area and exploring a random geographical area where the current species occurs. With variable Markov time, it explores low-degree nodes faster, connecting them stronger to other parts of the network. This has two effects. First, less diverse geographical nodes will be less likely to split into separate communities. If the low diversity is due to under-sampling, variable Markov time makes it harder to overfit on spurious patterns. Second, narrowly distributed species get a relatively stronger weight in pulling the geographical nodes where it exists together in a single community. Several studies have stressed the importance of narrowly distributed species for unveiling biogeographic patterns and evolutionary processes~\cite{quintero_global_2018}. Such species are usually also considered of higher conservation value~\cite{farooq_wege_2020}. These considerations suggest that the use of the variable Markov time algorithm may be more adequate for biological data sets of species occurrences. 

There is also a ``naturally occurring'' variable Markov time present in species occurrence networks with high proportions of endemic species.
When a random walker moves to an endemic species it is constrained to return to the same area it came from, slowing down network exploration. This effectively lowers the Markov time for some nodes more than others.

To evaluate the effect of variable Markov time, we partitioned the species occurrence network 1000 times and discarded the worst percentile. We then applied significance clustering~\cite{rosvall_mapping_2010} to find the community cores defined as the subset of nodes within the community that is partitioned together in at least 90 percent of the solutions. The other nodes' opacity indicates the fraction of times they are partitioned together with the community core. We draw a white boundary around each bioregion with varying opacity and width to indicate the fraction of times the two nodes on each side of the border are partitioned in different communities, where a wide opaque line means that the boundary is robust.

Infomap finds a hierarchical solution both with and without Variable Markov time, with bioregions nested within super-bioregions. However, at the super-level, many geographical cells are outside the community core, suggesting that the super-level boundaries are not robust across the world. At the next level, bioregions are more robust and variable Markov time typically keeps more regions together. For example, Europe and North Africa are in the red community, and the west of North America is in the light blue (Fig.~\ref{fig:bioregions}).

\section{Conclusion}
We have introduced variable Markov dynamics to identify non-clique-like community structures in a significantly broader range of scales than was previously possible. It does so by relaxing the constraint of the standard random walker to traverse the network with a constant speed to instead move faster on sparse structures. This flow model better matches the dynamics on many real-world networks that contain sparse functional structures with large diameters. For example, cars move faster on highways than in dense city road networks and neural information moves faster in the longer insulated neural pathways of white matter than in the denser subnetworks of grey matter that it connects. For species occurrence networks, variable Markov dynamics balances the tendency to visit species-rich areas and widespread species by giving more strength to the species found in less diverse areas and to narrowly distributed species.

It brings the advantages of High Dynamic Range (HDR) imaging to community detection. In the context of imaging, high dynamic range means large variation in light levels within a scene. To capture the details in dark areas, you need a relatively long exposure time, which will overexpose and blend all light areas. On the other hand, to capture the details in light areas, you need a low exposure time which collects too little light from dark areas to see the details therein. HDR techniques enable cameras to adapt exposure locally to the level of light to capture scenes with a higher dynamic range.

If we color a network with a lightness level proportional to the visit frequencies of a random walker, the Markov time parameter function as an exposure time for imaging, where variable Markov time increases the local exposure of dark areas without overexposing the light ones. This makes it possible to capture the large-scale pattern of flow with similar improvements as high dynamic range cameras made possible for imaging.

\section{Data availability}

The Electronic circuits network is available at {\small\url{https://www.weizmann.ac.il/mcb/UriAlon/download/collection-complex-networks}}.
The World Checklist of Vascular Plants (WCVP) is available at {\small\url{http://wcvp.science.kew.org/}}.
All other data are available upon request from the corresponding author.


\section{Funding}
D.E., J.S., and M.R. were supported by the Swedish Research Council (2016-00796). A.H.\ was supported by the Swedish Foundation for Strategic Research, Grant No. SB16-0089. A.A. acknowledges financial support from the Swedish Research Council (2019-05191), the Swedish Foundation for Strategic Environmental Research MISTRA (Project BioPath), and the Royal Botanic Gardens, Kew.

\section{Author contributions}

D.E.\ devised the study. D.E, J.S, and A.H. performed the experiments. 
A.A.\ suggested and interpreted the bioregionalization analysis.
All authors wrote, edited, and accepted the manuscript in its final form.

\section{Competing interests}

The authors declare no competing interests.

\medskip

\printbibliography

@article{fortunato_resolution_2007,
  title        = {Resolution limit in community detection},
  volume       = {104},
  pages        = {36--41},
  number       = {1},
  journaltitle = {Proc. Natl. Acad. Sci. USA},
  author       = {Fortunato, Santo and Barthélemy, Marc},
  year         = {2007}
}

@article{kheirkhahzadeh_efficient_2016,
  title        = {Efficient community detection of network flows for varying Markov times and bipartite networks},
  volume       = {93},
  pages        = {032309},
  number       = {3},
  journal = {Phys. Rev. E},
  author       = {Kheirkhahzadeh, Masoumeh and Lancichinetti, Andrea and Rosvall, Martin},
  year         = {2016}
}

@article{lancichinetti_community_2009,
  title        = {Community detection algorithms: A comparative analysis},
  volume       = {80},
  pages        = {056117},
  number       = {5},
journal = {Phys. Rev. E},
  author       = {Lancichinetti, Andrea and Fortunato, Santo},
  year         = {2009}
}

@article{schaub_markov_2012,
  title        = {Markov Dynamics as a Zooming Lens for Multiscale Community Detection: Non Clique-Like Communities and the Field-of-View Limit},
  volume       = {7},
  pages        = {e32210},
  number       = {2},
  journal = {{PLOS} {ONE}},
  author       = {Schaub, Michael T. and Delvenne, Jean-Charles and Yaliraki, Sophia N. and Barahona, Mauricio},
  year         = {2012}
}

@article{schaub_encoding_2012,
  title        = {Encoding dynamics for multiscale community detection: Markov time sweeping for the map equation},
  volume       = {86},
  pages        = {026112},
  number       = {2},
  journal = {Phys. Rev. E},
  author       = {Schaub, Michael T. and Lambiotte, Renaud and Barahona, Mauricio},
  year         = {2012}
}

@article{edler_infomap_2017,
  title        = {Infomap Bioregions: Interactive Mapping of Biogeographical Regions from Species Distributions},
  volume       = {66},
  pages        = {197--204},
  number       = {2},
  journaltitle = {Systematic Biology},
  author       = {Edler, Daniel and Guedes, Thaís and Zizka, Alexander and Rosvall, Martin and Antonelli, Alexandre},
  year         = {2017}
}

@article{rosvall_maps_2008,
  title        = {Maps of random walks on complex networks reveal community structure},
  volume       = {105},
  pages        = {1118--1123},
  number       = {4},
  journal = {Proc. Natl. Acad. Sci. USA},
  author       = {Rosvall, Martin and Bergstrom, Carl T.},
  year         = {2008}
}

@article{peixoto_parsimonious_2013,
  title        = {Parsimonious Module Inference in Large Networks},
  volume       = {110},
  pages        = {148701},
  number       = {14},
  journal = {Phys. Rev. Lett.},
  author       = {Peixoto, Tiago P.},
  year         = {2013}
}

@article{kawamoto_estimating_2015,
  title        = {Estimating the resolution limit of the map equation in community detection},
  volume       = {91},
  pages        = {012809},
  number       = {1},
  journal = {Phys. Rev. E},
  author       = {Kawamoto, Tatsuro and Rosvall, Martin},
  year         = {2015}
}

@report{page_pagerank_1999,
  title       = {The {PageRank} citation ranking: Bringing order to the web.},
  shorttitle  = {The {PageRank} citation ranking},
  institution = {Stanford {InfoLab}},
  author      = {Page, Lawrence and Brin, Sergey and Motwani, Rajeev and Winograd, Terry},
  date        = {1999}
}

@article{lambiotte_ranking_2012,
  title        = {Ranking and clustering of nodes in networks with smart teleportation},
  volume       = {85},
  pages        = {056107},
  number       = {5},journal = {Phys. Rev. E},
  author       = {Lambiotte, R. and Rosvall, M.},
  year         = {2012}
}

@article{rosvall_memory_2014,
  title        = {Memory in network flows and its effects on spreading dynamics and community detection},
  volume       = {5},
  pages        = {4630},
  number       = {1},
  journal = {Nat Commun},
  author       = {Rosvall, Martin and Esquivel, Alcides V. and Lancichinetti, Andrea and West, Jevin D. and Lambiotte, Renaud},
  year         = {2014}
}

@article{peixoto_merge-split_2020,
  title        = {Merge-split Markov chain Monte Carlo for community detection},
  volume       = {102},
  pages        = {012305},
  number       = {1},
  journal = {Phys. Rev. E},
  author       = {Peixoto, Tiago P.},
  year         = {2020}
}

@article{traag_louvain_2019,
  title        = {From Louvain to Leiden: guaranteeing well-connected communities},
  volume       = {9},
  pages        = {5233},
  number       = {1},
  journal = {Sci Rep},
  author       = {Traag, V. A. and Waltman, L. and van Eck, N. J.},
  year         = {2019}
}

@article{poldrack_long-term_2015,
  title        = {Long-term neural and physiological phenotyping of a single human},
  volume       = {6},
  pages        = {8885},
  number       = {1},
journal = {Nat Commun},
  author       = {Poldrack, Russell A. and Laumann, Timothy O. and Koyejo, Oluwasanmi and Gregory, Brenda and Hover, Ashleigh and Chen, Mei-Yen and Gorgolewski, Krzysztof J. and Luci, Jeffrey and Joo, Sung Jun and Boyd, Ryan L. and Hunicke-Smith, Scott and Simpson, Zack Booth and Caven, Thomas and Sochat, Vanessa and Shine, James M. and Gordon, Evan and Snyder, Abraham Z. and Adeyemo, Babatunde and Petersen, Steven E. and Glahn, David C. and Reese Mckay, D. and Curran, Joanne E. and Göring, Harald H. H. and Carless, Melanie A. and Blangero, John and Dougherty, Robert and Leemans, Alexander and Handwerker, Daniel A. and Frick, Laurie and Marcotte, Edward M. and Mumford, Jeanette A.},
  year         = {2015}
}

@article{govaerts_world_2021,
  title        = {The World Checklist of Vascular Plants, a continuously updated resource for exploring global plant diversity},
  volume       = {8},
  pages        = {215},
  number       = {1},
  journal = {Sci Data},
  author       = {Govaerts, Rafaël and Nic Lughadha, Eimear and Black, Nicholas and Turner, Robert and Paton, Alan},
  year         = {2021}
}

@book{brummitt_world_2001,
  title     = {World geographical scheme for recording plant distributions},
  volume    = {951},
  publisher = {International working group on taxonomic databases for plant sciences},
  author    = {Brummitt, Richard Kenneth and Pando, Francisco and Hollis, S. and Brummitt, N. A.},
  date      = {2001}
}

@article{tietje_global_2022,
  title        = {Global variation in diversification rate and species richness are unlinked in plants},
  volume       = {119},
  pages        = {e2120662119},
  number       = {27},
  journal = {Proc. Natl. Acad. Sci. USA},
  author       = {Tietje, Melanie and Antonelli, Alexandre and Baker, William J. and Govaerts, Rafaël and Smith, Stephen A. and Eiserhardt, Wolf L.},
  year         = {2022}
}

@article{quintero_global_2018,
  title        = {Global elevational diversity and diversification of birds},
  volume       = {555},
  pages        = {246--250},
  number       = {7695},
  journaltitle = {Nature},
  author       = {Quintero, Ignacio and Jetz, Walter},
  year         = {2018}
}

@article{farooq_wege_2020,
  title        = {{WEGE}: A new metric for ranking locations for biodiversity conservation},
  volume       = {26},
  pages        = {1456--1466},
  number       = {11},
  journaltitle = {Diversity and Distributions},
  author       = {Farooq, Harith and Azevedo, Josué and Belluardo, Francesco and Nanvonamuquitxo, Cristóvão and Bennett, Dominic and Moat, Justin and Soares, Amadeu and Faurby, Søren and Antonelli, Alexandre},
  year         = {2020}
}

@article{rosvall_mapping_2010,
  title        = {Mapping Change in Large Networks},
  volume       = {5},
  pages        = {e8694},
  number       = {1},
  journal = {{PLOS} {ONE}},
  author       = {Rosvall, Martin and Bergstrom, Carl T.},
  year         = {2010}
}

@software{Edler_The_MapEquation_software_2022,
author = {Edler, Daniel and Holmgren, Anton and Rosvall, Martin},
month = {10},
title = {{The MapEquation software package}},
url = {https://mapequation.org},
version = {2.6.1},
year = {2022}
}

@software{WCVP,
title = {{WCVP (2022). World Checklist of Vascular Plants, version 2.0. Facilitated by the Royal Botanic Gardens, Kew.}},
url = {http://wcvp.science.kew.org/},
year = {Retrieved 07 November 2022}
}

@ARTICLE{vinh2010ami,
  title = "{I}nformation {T}heoretic {M}easures for {C}lusterings {C}omparison: {V}ariants, {P}roperties, {N}ormalization and {C}orrection for {C}hance",
  author = "N. X. Vinh and J. Epps and J. Bailey",
  year = "2010",
  journal = "J. Mach. Learn. Res.",
  volume = "11",
  pages = "2837--2854",
}

@article{transport,
author = {Manlio De Domenico  and Albert Solé-Ribalta  and Sergio Gómez  and Alex Arenas },
title = {Navigability of interconnected networks under random failures},
journal = {Proc. Natl. Acad. Sci. USA},
volume = {111},
number = {23},
pages = {8351-8356},
year = {2014}
}

@article{elcircuits,
author = {Ron Milo  and Shalev Itzkovitz  and Nadav Kashtan  and Reuven Levitt  and Shai Shen-Orr  and Inbal Ayzenshtat  and Michal Sheffer  and Uri Alon },
title = {Superfamilies of Evolved and Designed Networks},
journal = {Science},
volume = {303},
number = {5663},
pages = {1538-1542},
year = {2004}
}

@article{streets,
  title = {Centrality measures in spatial networks of urban streets},
  author = {Crucitti, Paolo and Latora, Vito and Porta, Sergio},
  journal = {Phys. Rev. E},
  volume = {73},
  number = {3},
  pages = {036125},
  numpages = {5},
  year = {2006},
  publisher = {American Physical Society}
}

%

\renewcommand\thefigure{S\arabic{figure}}    
\setcounter{figure}{0}   

\section{Supplementary Information}

\begin{figure}[htbp]
  \centering
  \includegraphics[width=0.87\textwidth]{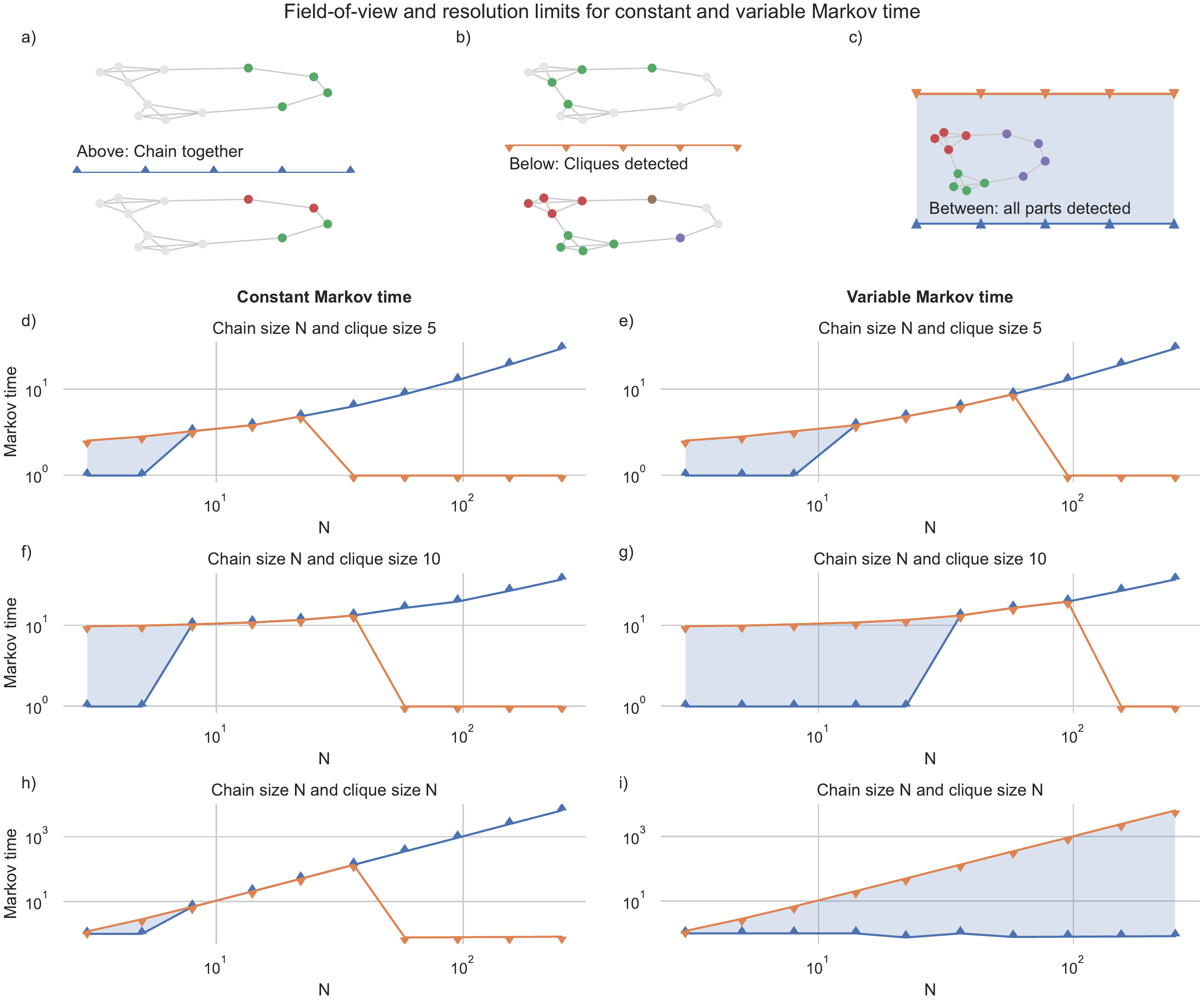}
  \caption{Field-of-view and resolution limits for constant and variable Markov time. A field-of-view limit is an upper scale on the size of a graph structure where the community detection method will over-partition larger structures. A resolution limit on the other hand is a lower scale on the size of a graph structure where the community detection method will under-partition smaller structures. The Markov time is a global resolution parameter that describes the step length of a random walker and can be increased from the default value of 1 to detect larger structures, but this also raises the resolution limit. The chain is kept together in one module in the space above the blue line, which marks the field-of-view limit in terms of a minimum Markov time needed to not break the chain (a). The cliques are detected below the orange line, which marks resolution limits (b). All three parts -- the two cliques and the chain -- are detected Between these limits (c). With constant Markov time, the chain is only detected below length 7 (d,f,h). With Variable Markov time, the field-of-view limit is pushed further away with increasing clique size (e, g), and disappears when the clique size grows the same as the chain (i).}
  \label{si-fig:fov}
\end{figure}

\clearpage

\subsection*{Interpolation between linear and logarithmic scaling using Tsallis entropy}

The Tsallis entropy for a probability distribution $P = \{p_1,p_2,\dots,p_n\}$ over $n$ categories is defined as
\begin{equation}
S_q(P) = \frac{k}{q - 1} \left( 1 - \sum_{i=1}^n p_i^q \right),
\end{equation}
where $q$ is a real parameter called entropic index and $k$ is a positive constant, hereafter assumed to be $1$.  In the limit as $q \to 1$ we recover the entropy
\begin{equation}
S_1(P) =  -\sum_{i=1}^n p_i \ln p_i
\end{equation}
which for a uniform distribution becomes $S_1(n) = \ln n$. To follow the convention used for the map equation to measure entropy in bits, the Tsallis entropy for a uniform distribution $p_i = 1/n$ is 
\begin{equation}
S_q(n) = \frac{n^{1-q} - 1}{(1-q)\ln 2}
\end{equation}
which gives $S_0(n) = (n - 1)/\ln2$ and $S_1(n) = log_2(n)$. To get an interpolator $f_q(n)$ from $f_{q=0}(n) = n$ to $f_{q=1}(n) = \log_2 n$, we define
\begin{equation}
f_q(n) = \frac{n^{1-q} - 1}{(1-q) \ln 2} c_s(q) + c_t(q)
\end{equation}
where $c_s(0) = \ln2, c_t(0) = 1$ and $c_s(1) = 1, c_t(1) = 0$.
To generalize for any real $q \geq 0$ with a smooth transition for the corrections at $q=1$ we adopt the simplest polynomial interpolator from zero to one on the domain $[0,1]$ with zero derivative at $q=1$, $s(q) = x(2 - x)$, and define the scale correction
\begin{equation}
    c_s(q) = \begin{cases} (1 - q(2-q)) \ln 2 + q(2 - q) \quad &\text{if }  0 \leq q < 1 \\
    1 &\text{if } q \geq 1
    \end{cases}
\end{equation}
and the translation correction
\begin{equation}
    c_t(q) = \begin{cases} 1 - q (2 - q) \quad &\text{if } 0 \leq q < 1 \\ 
    0 &\text{if } q \geq 1.
    \end{cases}
\end{equation}

\clearpage

\subsection*{Variable Markov time in Infomap}

We have implemented variable Markov time in the community detection algorithm Infomap version~2.6.1~\cite{Edler_The_MapEquation_software_2022}.
To enable this feature, use the command line argument \texttt{--variable-markov-time} and optionally \texttt{--variable-markov-damping d} where $d$ is a real number between 0 and~1.
Using the Infomap Python package, variable Markov time is enabled using the class constructor arguments:
\begin{lstlisting}[language=python]
from infomap import Infomap
im = Infomap(variable_markov_time=True, variable_markov_damping=d)
im.read_file("network.net")
im.run()
\end{lstlisting}


\end{document}